\documentclass[3p]{elsarticle}

\usepackage{lineno,hyperref}
\modulolinenumbers[5]

\journal{Journal of Magnetism and Magnetic Materials}

\usepackage[english]{babel}
\usepackage{epstopdf}
\usepackage{xcolor} 
\usepackage{soul} 
\usepackage{amsmath,amsfonts,mathrsfs,amsbsy,bm,babel,amssymb}
\usepackage{dcolumn}
\usepackage{bm}
\usepackage[utf8x]{inputenc}
\usepackage{float}
\usepackage{fixltx2e}
\usepackage{textcomp}
\usepackage[version=3]{mhchem}
\newcommand{\nH}{n_{\textsc{\tiny \emph{H}}}}
\newcommand{\SZ}{\langle S_Z \rangle}
\newcommand{\LZ}{\langle L_Z \rangle}
\newcommand{\TZ}{\langle T_Z \rangle}
\newcommand{\MZ}{\langle M_{total} \rangle}
\newcommand{\SZeff}{\langle S_Z^{\textsc{\tiny \emph{eff}}}\rangle}

\begin{document}

\begin{frontmatter}

\title{Magnetic properties of Dy nano-islands on graphene}
\author[ISU]{Nathaniel A. Anderson}
\author[ISU]{Qiang Zhang}
\author[ISU]{Myron Hupalo}
\author[Argonne]{Richard A. Rosenberg}
\author[Argonne]{John W. Freeland}
\author[ISU]{Michael C. Tringides}
\author[ISU]{David Vaknin}
\ead{vaknin@ameslab.gov}

\address[ISU]{Ames Laboratory and Department of Physics and Astronomy, Iowa State University, Ames, IA, 50011, USA}
\address[Argonne]{Advanced Photon Source, Argonne National Laboratory, Argonne, IL 60439}
\date{\today}

\begin{abstract}
We have determined the magnetic properties of epitaxially grown Dy islands on graphene/SiC(0001) that are passivated by a gold film (deposited in the ultra-high vacuum growth chamber) for {\it ex-situ} X-ray magnetic circular dichroism (XMCD).   Our sum-rule analysis of the Dy $M_{4,5}$ XMCD spectra at low temperatures ($T=15$ K) as a function of magnetic field assuming Dy$^{3+}$ (spin configuration $^6H_{15/2}$) indicate that the  projection of the magnetic moment along an applied magnetic field of 5 T is 3.5(3) $\mu_B$.  Temperature dependence of the magnetic moment (extracted from the $M_5$ XMCD spectra) shows an onset of a change in magnetic moment at about 175 K in proximity of  the transition from paramagnetic to helical magnetic structure at $T_{\rm H} =179$ K in bulk Dy.  No feature at the vicinity of the ferromagnetic transition of hcp bulk Dy at $T_{\rm c}$ = 88 K is observed. However, below $\sim$130 K,  the inverse magnetic moment (extracted from the XMCD) is linear in temperature as commonly expected from a paramagnetic system suggesting different behavior of Dy nano-island than bulk Dy. 
\end{abstract}

\begin{keyword}
Dysprosium\sep Nano-Islands\sep XMCD \sep STM
\end{keyword}

\end{frontmatter}


\section{Introduction}

Graphene has attracted extensive interest over the last decade due to its unique electronic properties and  its  potential application in microelectronics, catalysis, and spintronics.\cite{Vo-Van2011,Novoselov2004,CastroNeto2009}. To make it viably commercial,  it is necessary to understand its interactions with various metals either as dopants, as a substrate for functional nano-particles (i.e., nano-magnets), or for electrical contacts\cite{Karpan2007}. Metals have been used as dopants to control the Fermi level via charge transfer\cite{Gierz2008} but such transfer has to be sufficiently moderate to not  extensively degrade the innate graphene electronic properties. Growing magnetic metals (films or islands) on graphene are of particular interest  as  they can be used as spin filters\cite{Karpan2007} in spintronics or as high anisotropy nano-magnets\cite{Vo-Van2010}. There has been considerable progress in the growth and structural characterization of various metals on  graphene on 6H-SiC(0001)\cite{Liu2013}. Using STM in conjunction with DFT calculations, the growth morphology of Gd, Dy, Fe, Eu, and Pb as a function of temperature, flux, and coverage have been determined\cite{Hupalo2009,Liu2010,Hupalo2011,Hershberger2013}.  For instance, it has been found that Fe is anomalous as it shows morphologies that are not consistent with classical nucleation suggestive of the presence of long-range repulsive interactions. Our broader objective is to grow epitaxially magnetic metal films and islands of nanoscale dimensions and to determine their magnetization, magnetic domain distribution, and test for electron confinement effects, i.e., quantum size effects. 

Here, we report on X-ray magnetic circular dichroism (XMCD)  to determine the magnetic properties of Dy nano-islands grown on graphene. Bulk Dy is paramagnetic (PM) at room temperature and undergoes a first order magnetic transition from paramagnetic to incommensurate helical structure with a temperature dependent pitch at $T_H = 179$ K \cite{Wilkinson1961,Yu2015}. At $T_C \approx 88$ K Dy undergoes a first order phase transition from the helical structure to a ferromagnetic (FM) structure.  Recently, there has been renewed interest in the behavior of bulk Dy in magnetic field\cite{Yu2015}, as it has been reported that  in-plane applied magnetic field in the  temperature range $T_H \gtrsim T \gtrsim T_C$ induces field dependent {\it fan-like} intermediate phases \cite{Herz1978,Wakabayashi1997,Chernyshov2005}.  As discussed below, we show that the nano-islands grown in this study adopt the fcc(111) structure in contrast to the natural hcp structure of bulk Dy\cite{Hershberger2013}. 
We note that  in addition of the effect of crystal structural, the magnetic anisotropy can also depend on the size of the nano-crystals\cite{Kleibert2011}. It has been reported that hcp(1000) Dy islands  grown on W(110)  are ferromagnetic at $\approx$25 K\cite{Berbil-Bautista2007}.  However, the magnetization direction for islands less than $\sim80$ monolayers (ML) is in-plane instead  compared to out-of-plane as expected from bulk crystal. The ferromagnetism in the Dy islands on W(110) was studied by SP-STM and 6-fold degenerate magnetic domains were found to be larger than $\sim50$ nm. These sizes are much larger than the average size of the fcc Dy islands grown on graphene reported in the current study. Furthermore, recent XMCD studies of two Eu sub-monolayer (ML) phases $2\times2$   intercalated between graphene and Ir(111) have shown very different magnetic properties (despite the minute coverage difference) from those of bulk Eu\cite{Schumacher2014}. The $2\times2$ phase was found to be isotropically paramagnetic at 10 K with susceptibility 6 times lower than the bulk Eu susceptibility\cite{Schumacher2014}. The $\sqrt{3}\times\sqrt{3}$ phase is practically a ferromagnet with the easy axis normal to the plane and the in-plane magnetization being 20 times lower than the magnetization along the normal direction. Bulk Eu undergoes a first order PM-AFM transition at $T_N =88.4$ K\cite{Cohen1969}. In addition, previous experiments on isolated transition metal adatoms on graphene have shown lower magnetic moments than those in bulk environment \cite{Eelbo2013} as a result of strong hybridization with the carbon atoms. 

\begin{figure}[H] \centering \includegraphics [width = 3.4 in] {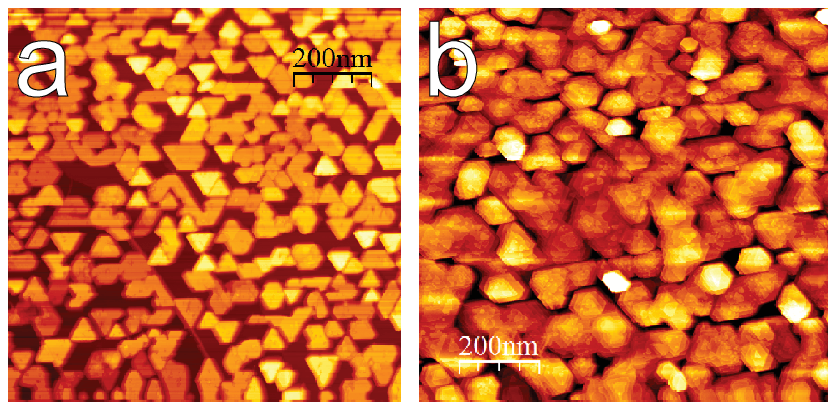}
\caption{(Color online) STM images of  (a) 9 ML ({\it thin}) and (b) 27 ML ({\it thick)} Dy nano-islands on graphene. After STM imaging and while still under ultra-high vacuum, samples were capped with 20 ML Au for use in this study.}
\label{fig:DySTM} 
\end{figure}
\section{Experimental Details}
X-ray magnetic circular dichroism (XMCD) measurements were performed at beamline 4-ID-C at the Advanced Photon Source in a chamber equipped with a high field ($<6$ T) split-coil, superconducting magnet. Detailed XMCD hysteresis loops were recorded at two X-ray helicities at the Dy $M_5$  edge (1295 eV) and $M_4$ edge (1333 eV). XMCD spectra were collected in helicity-switching mode at several temperatures in external magnetic fields applied parallel to the incident X-ray beam. The X-ray incident angle was fixed at $\approx 12^{\circ} \pm 2^{\circ}$ with respect to the sample surface and measurements were carried out by collecting the total electron yield (TEY), which was determined from the drain current to the sample.  XMCD data were collected by recording the helicity-dependent intensities, $\mu_+$  and $\mu_-$ , as a function of incident X-ray energy. The signals were normalized to the incident intensity (I0, monitor) by using the restoring current from a gold grid inserted into the X-ray beam path. Further details on the data analysis can be found in the Supplemental Information (SI).

The substrate used in the experiments was 6H-SiC(0001) purchased from Cree, Inc. The samples were graphitized in ultra-high vacuum (UHV, P $\approx1\cdot10^{-10}$ Torr) by direct current heating of the sample to ≈ 1300 C measured with an infrared pyrometer. The crystalline Dy islands were grown on the graphene surface by depositing Dy using a molecular beam source at a flux rate of 0.1 - 0.2  monolayers (ML)/min\cite{Hupalo2009}. The Dy source was degassed during the bake-out for several hours, so during deposition the pressure remained below $1.6\cdot10^{-10}$ Torr. Both samples were capped with 20 layers of Au, while still under ultra-high vacuum, for protection against oxidation during sample transport (in air) for the XMCD experiments. The number of ML of Dy is determined by finding the integrated island volume within a given area after correcting for the usual convolution of tip effects. For this experiment, the samples had a deposited amount of $\theta=9$ ML and $\theta=27$ ML (Fig. \ref{fig:DySTM}(a) and \ref{fig:DySTM}(b), respectively). For $\theta=9$ ML the average height is 3 nm with a standard deviation 1 nm while the average area is $2\cdot10^{3}$ nm$^2$. The distributions for the thick sample, $\theta=27$ ML, are 8 nm for the height and 6000 nm$^2$ for the area with a similar uncertainty.

As shown in Fig. \ref{fig:DySTM}, the grown islands have triangular shapes, unusual for hcp(0001) crystals which exhibit hexagonal symmetry with six facet planes\cite{Hershberger2013}. On the other hand, the six facet planes surrounding an fcc(111) island are not equivalent, they alternate between (111) and (100)  planes . If the planes have different surface energies or if the kinetic barriers controlling the attachment of atoms to the facets are different and one of the two orientations is favored, then it is possible to grow triangular islands. Thus, the triangular shapes suggest the islands are fcc(111). Further support of this conclusion is the stacking sequence of Dy islands with incomplete tops\cite{Hershberger2013}. These islands were intentionally grown in stepwise coverage deposition procedures, i.e. the same total amount was deposited not in a single deposition but in several smaller depositions. With stepwise deposition, the islands grow with larger bases and successive layers nucleating in a {\it wedding cake}-like morphology and remain incomplete. For fcc islands with ABCABC  stacking the successive incomplete planes should point in the same direction; while  for hcp(0001) islands with ABABAB stacking the incomplete planes should alternate in direction. As shown in Ref. \cite{Hershberger2013} successive planes on the incomplete Dy islands point to the same direction confirming that the islands are fcc(111).

\section{Results and Discussion}
\begin{figure}[H] \centering \includegraphics [width =3.5 in] {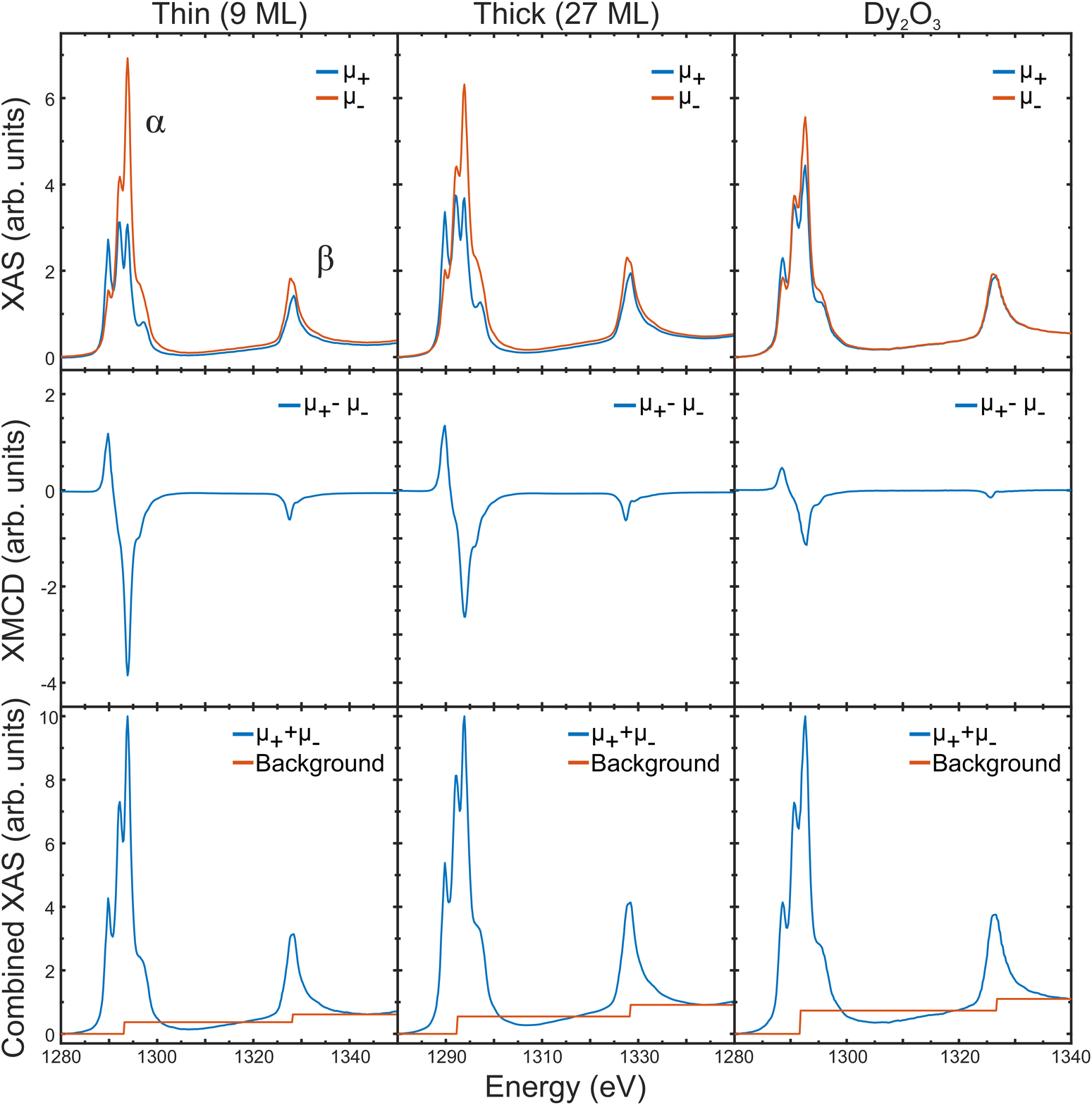} 
\caption{(Color online) The $\mu_+$ and $\mu_-$ XAS,  XMCD, and combined XAS from the {\it thin} (9 ML) and {\it thick} (27 ML) dysprosium nano-islands on graphene, and from polycrystalline Dy$_2$O$_3$ as indicated, at $B = 5$ T at $T = 15$ K. The steps are used for background subtraction when determining the integral $r=\int_{\alpha+\beta}(\mu_++\mu_-)$. The $\alpha$ and $\beta$ regions correspond to the $M_5$ and $M_4$ spectra, respectively.}
\label{fig:XAS/XMCD/aveXAS} 
\end{figure}
Fig. \ref{fig:XAS/XMCD/aveXAS} shows the XAS at the Dy $M_4$ and $M_5$ edges for the two circularly polarized X-ray beams at $T =15$ K and $B = 5$ T  for the {\it thin} and {\it thick} samples (Fig. \ref{fig:DySTM}) and for \ce{Dy2O3}, for comparison (Dy$_2$O$_3$ is paramagnetic to low temperature\cite{Chen2011}).  The difference between the two circularly polarized XAS ($\mu_+$ and  $\mu_-$) shows two prominent XMCD signals. The normalized XAS spectra of the three samples  in Fig. \ref{fig:XAS/XMCD/aveXAS} are consistent with and practically indistinguishable from bulk Dy metal.\cite{Thole1985} This, to our understanding, is due to the fact that the transitions correspond to core level electrons that are insignificantly affected by the oxidative state of Dy in ionic or metallic environments. Even though the XMCD signals also look similar, those of the Dy-islands are significantly stronger than that of  \ce{Dy2O3} (see more details on the magnetic properties of \ce{Dy2O3} in the SI), qualitatively indicating a larger average magnetic moment for the nano-islands. We also note that the XMCD signal of the thick sample is slightly reduced compared to the thin sample.  This may be due to slight morphology differences between the two samples or due to inadvertent partial oxidation suggesting that the Au capping does not perfectly passivate the samples.
To quantitatively determine the effective magnetic moment and the electronic configuration we use  sum rules derived by Carra \emph{et. al}\cite{Carra1993}:
\begin{equation}
\frac{p+q}{\frac{3}{2}r}=\frac{1}{2}\frac{\ell(\ell+1)+2-c(c+1)}{\ell(\ell+1)\nH}\LZ
\end{equation}
\begin{equation}
\frac{p-(\frac{c+1}{c})q}{\frac{3}{2}r}=\frac{\ell(\ell+1)-2-c(c+1)}{3c\nH}\SZ+\frac{\ell(\ell+1)[\ell(\ell+1)+2c(c+1)+4]-3(c-1)^2(c+2)^2}{6c\ell (l+1)\nH}\TZ
\end{equation}
where $\ell$ is the angular momentum quantum number of valence shell, $c$ is the angular momentum quantum number of core shell, $\nH$ is the number of electron holes in the valence shell,  $p=\int_\alpha\mu_+-\mu_-$, $q=\int_\beta\mu_+-\mu_-$, and {$r=\int_{\alpha+\beta}(\mu_++\mu_-)$. For Dy, the integration over $\alpha$ corresponds to the $M_5$ signal (energy range $1280 - 1305$ eV) and $\beta$ corresponds to the $M_4$ peak  ($1323 - 1348$ eV). The integrations of $p$, $q$, and $r$, are calculated using the trapezoidal rule\cite{Press1992} for each energy bin. The combined XAS, $r$, is background subtracted as described in the SI. Assuming $\text{Dy}^{3+}$, the angular momentum quantum number of the valence and core shells are $\ell=3$ and  $c=2$, respectively, and the number of electron holes in the valence shell is $\nH=5$. This significantly  simplifies the sum rules for $\text{Dy}^{3+}$ so that 
\begin{equation}
\langle L_Z \rangle=\frac{2(p+q)}{r}\nH 
\end{equation}
 and 
 \begin{equation}
\SZeff \equiv \SZ+3\TZ=\frac{2p-3q}{2r}\nH
 \end{equation}
 Using the relationship of $\frac{\SZ}{\LZ}$ given by Equation 18 of Collins \emph{et. al}\cite{Collins1995} and Hund's rules with $J=\frac{15}{2}$, $L=5$, and $S=\frac{5}{2}$, we arrive at $\frac{\SZ}{\LZ}=\frac{1}{2}$. This enables derivation of $\TZ$ in terms of $q$ and $r$ such that $\TZ=-\frac{5q}{6r}\nH$, and also simplifies the extraction of the total moment 
\begin{equation}
\MZ=\LZ+2\SZ=2\LZ
\end{equation}

\begin{figure}[H] \centering \includegraphics [width = 3.2 in] {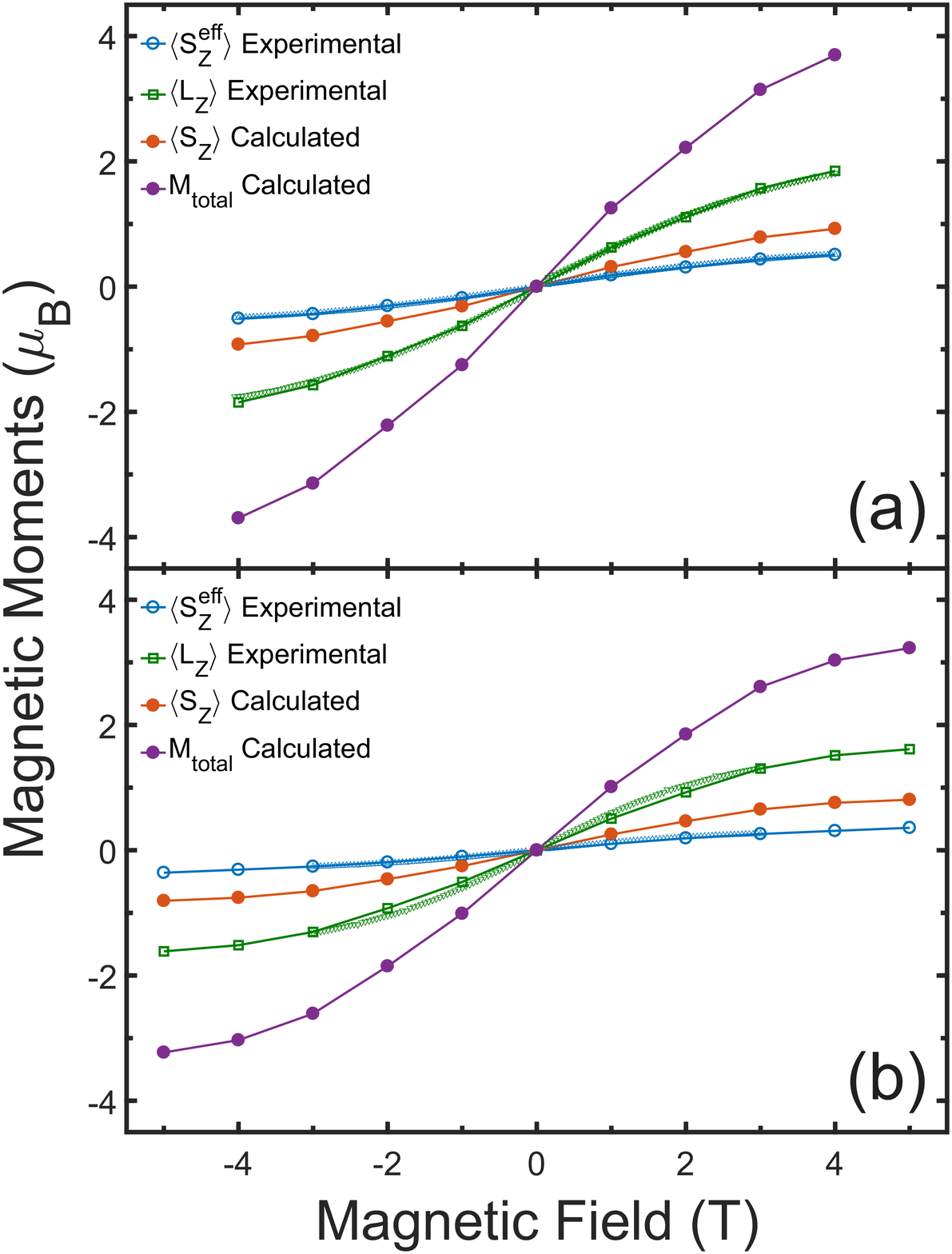}
\caption{(Color online) The different contributions to the total magnetic moment $\MZ$ versus magnetic field of (a) {\it thin} and (b) {\it thick} dysprosium layer on graphene at $T=15$ K. The thicker solid lines are in fact densely packed points corresponding to the $\LZ$ and $\SZeff$ (green and blue, respectively) measured with the fine B-field as detailed in the text.}
\label{fig:DyBfieldt} 
\end{figure}

Figure \ref{fig:DyBfieldt} shows the average spin and angular momentum components of the total moment, $\MZ$, as a function of the applied field at $T=15$ K, obtained by using the above sum rules for both the {\it thin} and {\it thick} samples (see SI for more details on data processing). This contrasts with reports on Dy films on W(110) which exhibit ferromagnetic order according to spin-polarized scanning tunneling microscopy and spectroscopy\cite{Berbil-Bautista2007}.  Applying similar analysis to the Dy$_2$O$_3$ polycrystalline sample yields the curve shown in Fig. \ref{fig:DyXMCDmom} with an effective total moment that is significantly smaller than those of the Dy nano-islands. The fine $B$ field dependence showing no hysteresis (Fig. \ref{fig:DyBfieldt}) is obtained by measuring the XMCD at a fixed photon energy (1293 eV for {\it thin}, 1295.5 eV for {\it thick}), where it is prominent and by scaling the data to the sum-rule-derived $\LZ$.  Examining the different components of the moment (i.e. $\LZ$, $\TZ$, $\SZeff$, etc.) we find that they all behave similarly with respect to the applied magnetic field up to a scale factor.  In particular, we find that the integral $q$ over the $M_4$ edge is proportional to $\TZ$ and thus, up to a scale factor, adequately represents the total magnetic moment, $\MZ$. It is therefore justified to use the normalized XMCD over the $M_4$ edge (up to a scaled factor) to monitor the temperature dependence of the magnetic moment in Dy islands.

\begin{figure}[H] \centering \includegraphics [width = 3.2 in] {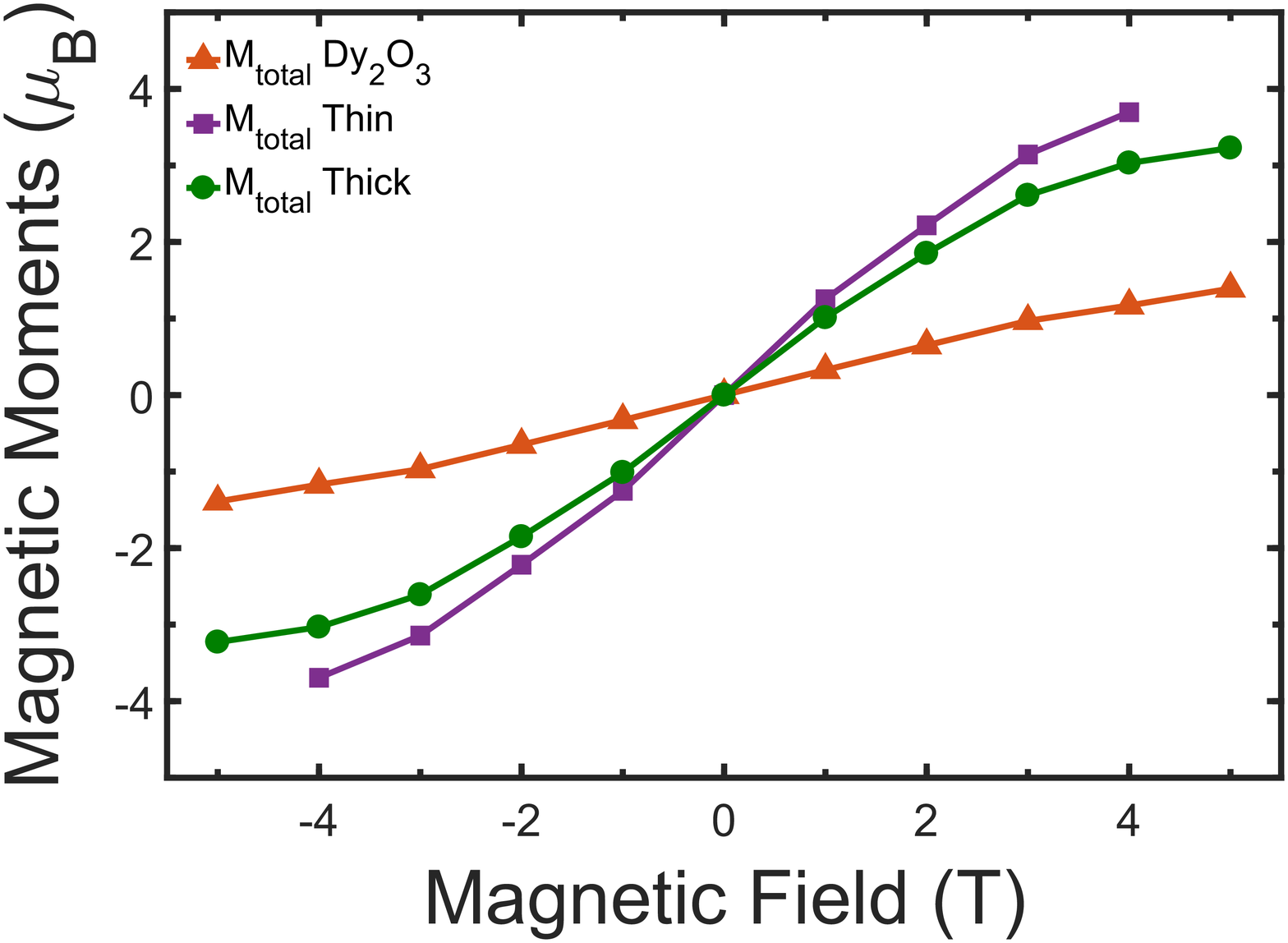}
\caption{(Color online) The total magnetic moments, $\MZ$,versus magnetic field of \ce{Dy2O3} powder (triangle), {\it thin} Dy islands (square), and {\it thick} Dy islands (circle). All scans were conducted at $T=15$ K. The \ce{Dy2O3} has a total moment substantially lower than that of the island samples. The small reduction of the {\it thick} sample compared to the {\it thin} could be due to partial oxidation.}
\label{fig:DyXMCDmom} 
\end{figure}

\begin{figure}[H] \centering \includegraphics [width = 2.7 in] {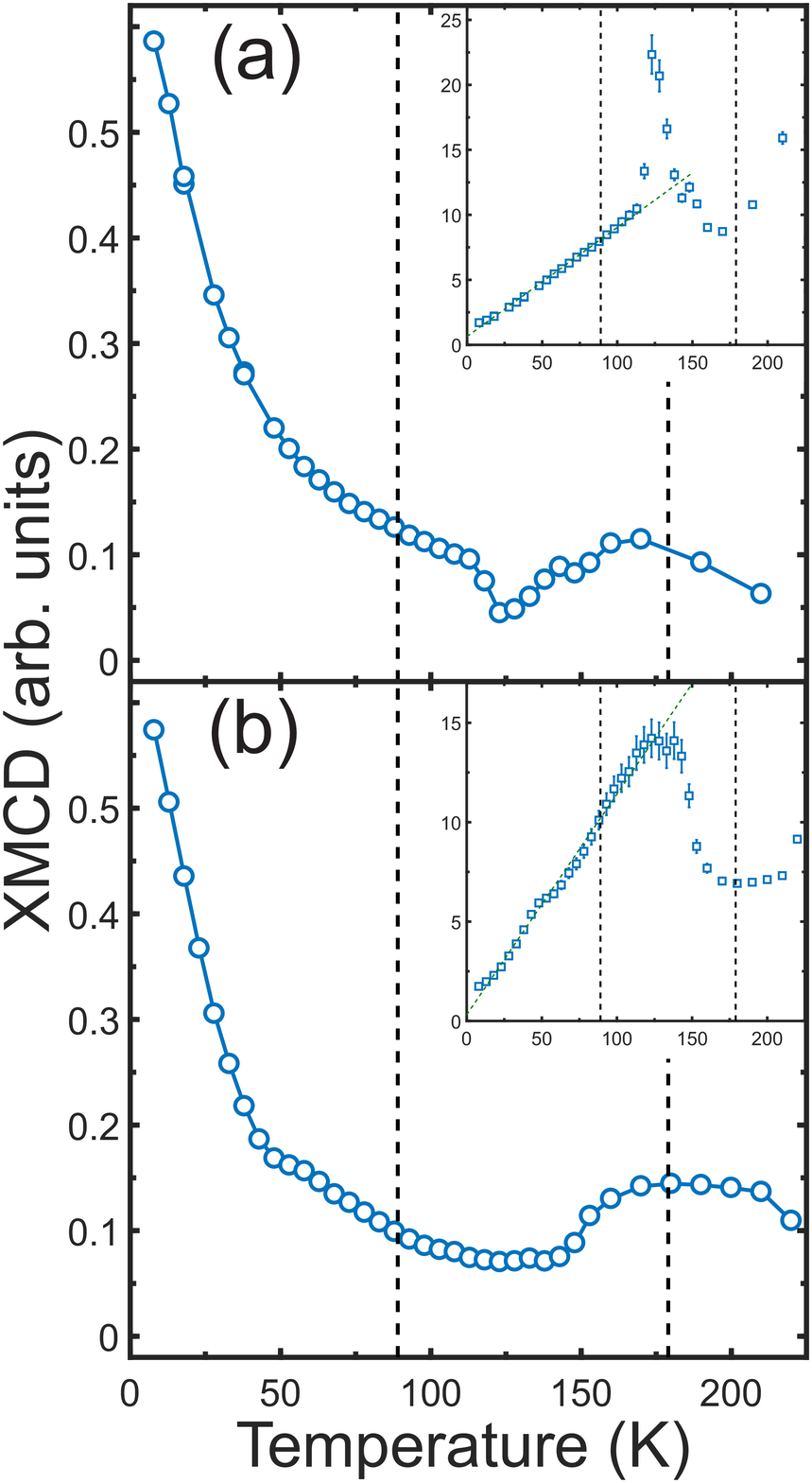}
\caption{(Color online) The temperature dependence of the XMCD and 1/XMCD (insert) of the integrated intensity at the $M_4$ peak for (a) {\it thin} and (b) {\it thick} Dy on graphene at $B=3$ T. Dashed lines indicate the transition temperatures $T_H=179$ and $T_C =88$  to helical and ferromagnetic phases, respectively, for bulk Dy\cite{Yu2015}.}
\label{fig:DyTemp} 
\end{figure}

Fig. \ref{fig:DyTemp} shows the temperature dependence of the  integrated XMCD over the $M_4$ spectra measured at $B=3 $ T, (namely, $q=\int_\beta\mu_+-\mu_- $) which is proportional to $\MZ$ by a scale factor. Also displayed as insets are the inverse integrated XMCD, $1/q$, as function of temperature.   Broad peaks in $q$ (or uptake in $1/q$) at around $T \sim 170$ K seem to coincide with a known transition of bulk Dy from the paramagnetic state to helical structure at $T_H = 179$ K \cite{Wilkinson1961,Yu2015}.  We note that the temperature evolution of the XMCD of the two samples (9 and 27 MLs) and as a function of magnetic field are very similar despite the fact that their morphology seems to be different (see Fig.\ \ref{fig:DySTM}).  Measurements to determine a blocking temperature superparamagnetism\cite{Liu2011} at a very low magnetic field ($B=0.01$ T) did not yield adequate XMCD signals (data not shown).

\section{Conclusions}
XMCD has been used to determine the magnetic properties of Dy nano-islands grown on graphene and covered with a Au film for passivation (Au capping performed under ultra high vacuum in the growth chamber).  We have studied Dy triangular-shaped islands of two different heights (9 and 27 ML) on epitaxial graphene and determined the contributions of the angular momentum $\LZ$, the spin $\SZ$ to the total magnetic moment, $\MZ$, where the dipolar term $\TZ$ is  simple for the $^6H_{15/2}$ configuration of Dy$^{3+}$, and it can be extracted from the $M_4$ edge via the XMCD data and sum rules.  We find that the total moment in the nano-islands is not saturated at $H = 5$ T at temperatures below which bulk-Dy is ferromagnetic  ($T_{\rm c} = 88$K).  We show that the temperature dependence of the average total magnetic moment can be reliably extracted from the $M_4$ spectra alone, showing that the XMCD exhibits anomalies at high temperatures that may be associated with  the known paramagnetic helical transition in bulk  Dy ($T_H = 189$ K).  However  below $\approx 130$ K the inverse magnetic moment exhibits linear temperature dependence  as commonly seen in paramagnetic systems with no obvious features in the vicinity of the helical-to-ferromagnetic transition in bulk Dy at $T_{\rm c} = 88$ K.  This behavior may be due to the morphology and size of the nano-structures with easy axis (111) normal to the substrate and to the role of graphene might play in the growth process. It is also known that even for uniformly grown films of Fe on Au(111), XMCD experiments show that the spin moment depends on film thickness and therefore more pronounced changes are expected for the nano-islands compared to their bulk parent materials\cite{Bayreuther1993}.  This behavior may also be due to partial  chemical conversion of the sample. In this regard, we note that the $M_4$ and $M_5$ spectra are due to core levels of Dy and therefore difficult to distinguish between elemental Dy metal from that of ionic Dy in a compound, and therefore we cannot rule out the possibility that some oxidation through the Au capping occurs.
\section{Acknowledgments}
Research at Ames Laboratory is supported by the US Department of Energy, Office 
of Basic Energy Sciences, Division of Materials Sciences and Engineering under 
Contract No. DE-AC02-07CH11358.  Use of synchrotron radiation at the Advance Photon Source, Argonne National Laboratories was supported by U.S. DOE under contract No. DE-AC02-06CH11357.

\section{Appendix}
\subsection{Derivations}
We can derive sum rules for $\TZ$ from the sum rules of $\LZ$, $\SZeff$, and $\frac{\SZ}{\LZ}=\frac{1}{2}$:
\begin{align} \label{eq:TzSum}
\SZ+3\TZ		&	=\frac{2p-3q}{2r}\nH \Rightarrow &\phantom{(}  				\notag\\
\TZ			&	=\frac{1}{3}\left(\frac{2p-3q}{2r}\nH-\SZ\right) 					\notag\\
\phantom{\TZ}	&	=\frac{2p-3q}{6r}\nH-\frac{1}{6}\LZ						 	\notag\\
\phantom{\TZ}	&	=\frac{2p-3q}{6r}\nH-\frac{2p+2q}{6r}\nH					\notag\\
\phantom{\TZ }	&	=-\frac{5q}{6r}\nH									
\end{align}

\bibliography{Dy-MMM_NAA.bbl}

\end{document}